\documentclass[10pt, conference, twocolumn]{IEEEtran}                                                  

\IEEEoverridecommandlockouts                              

\title{Secure Lossless Compression with Side Information}

\author{%
  \authorblockN{Deniz G\"{u}nd\"{u}z\authorrefmark{1}\authorrefmark{2},
    Elza Erkip\authorrefmark{1}\authorrefmark{3},
    H. Vincent Poor\authorrefmark{1}
  }\\
  \authorblockA{%
    \authorrefmark{1}Dept.\ of Electrical Engineering,
                     Princeton University, Princeton, NJ, 08544\\
  }
    \authorblockA{%
    \authorrefmark{2}Dept.\ of Electrical Engineering,
                     Stanford University, Stanford, CA, 94305\\
  }
  \authorblockA{%
    \authorrefmark{3}Dept.\ of Electrical and Computer Engineering,
                     Polytechnic University, Brooklyn, NY, 11201 \\
  }
  Email: \authorrefmark{1}\{dgunduz, poor\}@princeton.edu,
  \authorrefmark{3}elza@poly.edu
  \thanks{This research was supported in part by the US National Science Foundation under Grants  CCF-04-30885, CCF-06-35177, CCF-07-28208, and CNS-06-25637.}
}

\date{September, 2006}
\usepackage{amsfonts}
\usepackage{amssymb}
\usepackage{psfrag}
\usepackage{amsmath}
\usepackage{amscd}
\usepackage{graphicx}

\graphicspath{{fig/}{num/}}

\newtheorem{thm}{Theorem}[section]
\newtheorem{cor}[thm]{Corollary}

\newtheorem{defn}{Definition}[section]

\begin{document}
\maketitle \thispagestyle{empty}
\pagestyle{empty}

\begin{abstract}
Secure data compression in the presence of side information at both a legitimate receiver and an eavesdropper is explored. A noise-free, limited rate link between the source and the receiver, whose output can be perfectly observed by the eavesdropper, is assumed. As opposed to the wiretap channel model, in which secure communication can be established by exploiting the noise in the channel, here the existence of side information at the receiver is used. Both coded and uncoded side information are considered. In the coded side information scenario, inner and outer bounds on the compression-equivocation rate region are given. In the uncoded side information scenario, the availability of the legitimate receiver's and the eavesdropper's side information at the encoder is considered, and the compression-equivocation rate region is characterized for these cases. It is shown that the side information at the encoder can increase the equivocation rate at the eavesdropper. Hence, the side information at the encoder is shown to be useful in terms of security; this is in contrast with the pure lossless data compression case where side information at the encoder would not help.
\end{abstract}

\section{Introduction}
Consider a sensor network in which multiple sensors observe an underlying phenomenon that needs to be reconstructed at an access point. While some sensors might have secure (possibly wired) connections to the access point, others might be transmitting over the wireless medium, which can be accessed by an adversary trying to obtain information about the underlying phenomenon. Furthermore, this adversary might have its own observation of the main source. Our goal is to explore the security issues in this sensor network scenario. Our model is a simplified version of the general problem, in which we assume a single sensor (Alice) having direct access to the underlying source that needs to be transmitted to the access point (Bob) reliably and securely. Furthermore, we assume an idealized noise-free channel whose output can also be observed by the eavesdropper (Eve). 

\begin{figure}
\centering
\includegraphics[width=2.7in]{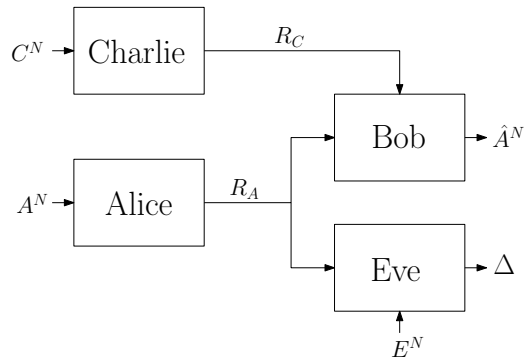}
\caption{Side information of Bob is provided by Charlie who has access to his own correlated side information.}
\label{f:codedSI}
\end{figure}

If no side information is available to Bob, then we cannot achieve any level of security. However, if we assume the existence of a nearby sensor (Charlie) having access to correlated side information about Alice's source and a secure limited-rate link to Bob, this sensor might enable secure transmission of Alice's source using its own secure link (see Fig. \ref{f:codedSI}). Our goal is to characterize the capacities of error-free communication links from Alice and Charlie to Bob such that Alice's information can be reliably transmitted to Bob, while keeping Eve's information about the source limited.

Secure communication over noisy channels in the presence of a wiretapper has recently attracted considerable interest. Information theoretic security in this context is defined through the equivocation rate at the wiretapper, which can be roughly defined as the uncertainty of the wiretapper about the message after observing the channel output. In his pioneering work \cite{Wyner}, Wyner introduced the wire-tap channel, and showed that it is possible to transmit at a positive rate with perfect secrecy, assuming the wiretapper's channel is physically degraded with respect to the receiver. Later, Wyner's analysis is extended to more general broadcast channels in \cite{Csiszar_Korner}, which characterizes the capacity-equivocation rate region. Various extensions of the wiretap channel model to multiuser scenarios and fading channels have recently been investigated \cite{Ender, Liang, Ruoheng}.

In the wiretap channel model, the potential for secure communication arises from the fact that the intended receiver has a better quality communication channel than the wiretapper \cite{Csiszar_Korner}. In our model, since the communication channels are not noisy, the techniques of \cite{Csiszar_Korner} do not apply; however, it is still possible to achieve security when Bob has higher quality side information than Eve as in \cite{Merhav, Ramc}. In  \cite{Merhav}, Merhav proved a source-channel separation theorem for the wiretap channel assuming both the channel and the side information of the wiretapper are physically degraded.  Recently, Prabhakaran and Ramchandran \cite{Ramc} consider the arbitrarily correlated side information case focusing only on the leakage rate to the eavesdropper. They find the minimum leakage rate, and through an example, argue that the availability of Bob's side information to Alice might increase Eve's uncertainty about Alice's source. Secure compression of two correlated sources is considered in \cite{Kundur}, where the eavesdropper has access to only one of the compressed bit streams. Our work is also closely related to the secret key capacity model of \cite{Ahlswede_Csiszar, Maurer}, where correlated sources are used for secure key generation. However, our goal here is not to generate a secret key among Alice and Bob. Instead, we wish to communicate Alice's source to Bob securely.

In this paper, we first consider the case in which the side information of Bob is provided by Charlie over a noise-free secure channel. After giving inner and outer bounds for the set of achievable compression-equivocation rates for this setup, we focus on the case in which Charlie-Bob link has enough capacity for Bob to obtain Charlie's side information losslessly. For this scenario, which also corresponds to uncoded side information, we consider cases in which either or both Bob's and Eve's side information may be available to Alice. We show that, in the secure compression model, as opposed to the usual lossless compression where side information at the encoder does not improve the performance, the availability of side information to Alice has the potential of improving the secrecy performance. We generalize the characterization of the achievable compression and equivocation rates to all the side information cases and provide illustrative examples.

\section{System Model}
We assume that Alice has access to an $N$-length source sequence $A^N$, which she wants to transmit to Bob reliably over a noise-free, finite capacity channel. Alice's transmission will also be perfectly received by an eavesdropper called Eve. We assume that Eve has her own correlated side information $E^N$. On the other hand, a helper, called Charlie, has access to correlated side information $C^N$ and a limited rate secure channel to Bob (see Fig. \ref{f:codedSI}). We model $A^N$, $C^N$, and $E^N$ as being generated independent and identically distributed (i.i.d.) according to the joint probability distribution $p_{A,C,E}(a,c,e)$ over the finite alphabet $\mathcal{A} \times \mathcal{C} \times \mathcal{E}$. While Alice wants to transmit her source reliably to Bob, she also wants to maximize the equivocation at Eve, which represents the uncertainty of Eve about $A^N$ after receiving Alice's transmission and combining with her (Eve's) own side information $E^N$.


An $(R_A,R_C,N)$ code for secure source compression in this setup is composed of an encoding function at Alice\footnote{To keep the presentation simple, here we assume deterministic coding, but similar to \cite{Ahlswede_Csiszar}, randomized coding can be considered by assuming that Alice, Bob and Charlie initially generate independent random variables and keep the rest of the coding scheme deterministic. Proofs would follow similarly.}, $f_A: \mathcal{A}^N \rightarrow \{1,2,\ldots, 2^{NR_A} \}$, an encoding function at Charlie, $f_C: \mathcal{C}^N \rightarrow \{1,2,\ldots, 2^{NR_C} \}$, and a decoding function at Bob, $g^N: \{1,2,\ldots, 2^{NR_A} \} \times \{1,2,\ldots, 2^{NR_C} \} \rightarrow \mathcal{A}^N$.

The equivocation rate of this code is defined as
\begin{eqnarray}
\frac{1}{N} H(A^N|f_A(A^N), E^N),
\end{eqnarray}
and the error probability of the code has the usual definition:
\begin{eqnarray}
P_e^{N} = P(g(f_A(A^N), f_C(C^N)) \neq A^N ).
\end{eqnarray}

\begin{defn}
We say that $(R_A, R_C, \Delta)$ is \emph{achievable} if, for any $\epsilon>0$, there exist an $(R_A,R_C,N)$ code such that $H(A^N|f_A(A^N), E^N) \geq N \Delta$ and $P_e^{N} < \epsilon$.
\end{defn}

\section{Coded and Uncoded Side Information at Bob}\label{s:no_enc_SI}

In this section, we give inner and outer bounds to the set of all achievable $(R_A,R_C,\Delta)$ triplets. In general, these bounds do not match.

\begin{thm}\label{t:codedSI}
For the setup above, $(R_A, R_C, \Delta)$ is achievable if,
\begin{eqnarray}
  R_A &\geq& H(A|V), \label{rateA}\\
  R_C &\geq& I(C;V),\label{rateC}\\
  \Delta & \leq & \max \{I(A;V|U) - I(A;E|U)\}, {~\rm and}  \label{rateD} \\
  R_A + \Delta & \geq & H(A|E), \label{rateT}
\end{eqnarray}
where we maximize over auxiliary random variables $V$ and $U$ that come from the joint distribution $p(a,c,e,u,v)= p(a,c,e)p(u|a)p(v|c)$ with $|\mathcal{U}| \leq |\mathcal{A}|+1$ and $|\mathcal{V}| \leq |\mathcal{C}|+2$.

Conversely, if $(R_A, R_C, \Delta)$ is achievable, then (\ref{rateA})-(\ref{rateT}) hold for some auxiliary random variables $V$ and $U$ for which $V-C-(A,E)$ and $U-A-(C,E)$ form Markov chains.
\end{thm}
\begin{proof}
The proof is given in Appendix \ref{App1}.
\end{proof}

We can consider this problem to be a generalization of source coding with coded side information \cite{Wyner_SI}, where we have the security constraint in addition to lossless compression. In the achievability of the inner bound given in Appendix \ref{App1}, Alice's encoder, instead of directly binning its observation with respect to the coded side information at Bob, uses an auxiliary codebook generated by $U$ to send her observation and creates higher equivocation at Eve. This auxiliary codebook generation resembles lossy source coding with coded side information \cite{Berger} for which the single letter characterization of the rate region remains to be an open problem. Similar to the inner and outer bounds for that problem \cite{Tung}, our inner and outer bounds differ in the joint distribution of the auxiliary random variables.

\begin{figure}
\centering
\includegraphics[width=2.5in]{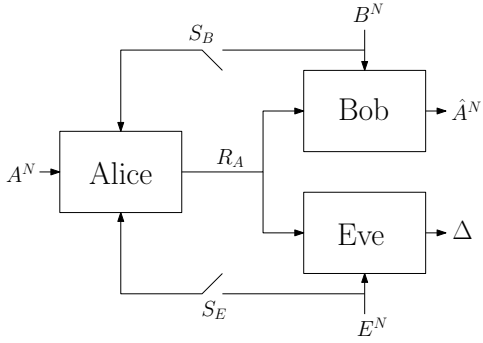}
\caption{Uncoded side information at Bob. The states of switches $S_B$ and $S_E$ model different scenarios in terms of the side information at the encoder.}
\label{f:uncodedSI}
\end{figure}

A special case of the above theorem is obtained when we assume that $R_C \geq H(C)$, that is, the side information $C^N$ of Charlie can be recovered by Bob with an arbitrarily small probability of error. In this scenario, in order to keep the presentation simple, we can assume that a side information sequence $B^N$ is available directly to Bob where $B^N=C^N$ with high probability (see Fig. \ref{f:uncodedSI} with both switches open). For this uncoded side information case, the decoding function at Bob is replaced by $g^N:\{1,2,\ldots, 2^{NR_A} \} \times \mathcal{B}^N  \rightarrow \mathcal{A}^N$. The achievability is now defined similarly, for an $(R_A, \Delta)$ pair.

We have the following corollary which follows from Theorem \ref{t:codedSI}. The proof of this special case (assuming no rate limitations between Alice and Bob) is also given in \cite{Ramc}.

\begin{cor}\label{c:uncodedSI}
For uncoded side information $B^N$ at Bob, $(R_A, \Delta)$ is an achievable rate-equivocation pair if and only if,
\begin{eqnarray}
  R_A &\geq& H(A|B), {~\rm and} \label{uncodedA} \\
  \Delta & \leq & \max \{I(A;B|U) - I(A;E|U)\},  \label{uncodedD}
\end{eqnarray}
where we maximize over auxiliary random variables $U$ such that $U-A-(B,E)$ form a Markov chain and $|\mathcal{U}| \leq |\mathcal{A}|+1$.
\end{cor}

While Corollary \ref{c:uncodedSI} requires an auxiliary codebook generated by $U$ in the general case to conceal the source from the eavesdropper, it is sometimes possible that the ordinary Slepian-Wolf binning achieves the highest possible security in terms of equivocation, i.e., (\ref{uncodedD}) is maximized by a constant $U$. Some definitions are in order.

\begin{defn}
We say that the side information $B$ is \textit{less noisy} than the side information $E$ if
\begin{eqnarray}
I(U;E) \leq I(U;B)
\end{eqnarray}
for every probability distribution of the form $p(a,b,e,u) = p(a,b,e)p(u|a)$.
\end{defn}

\begin{defn}
Side information $E$ is said to be \textit{physically degraded} with respect to $B$ if, $A-B-E$ form a Markov chain. We say $E$ is \textit{stochastically degraded} with respect to $B$ if, there exists a joint probability distribution $p_{A\tilde{B}\tilde{E}}$ such that $p_{A\tilde{B}}= p_{AB}$, $p_{A\tilde{E}}=p_{AE}$, and $A-\tilde{B}-\tilde{E}$ is a Markov chain.
\end{defn}

The \textit{less noisy} condition is strictly weaker than the \emph{stochastically degraded} condition \cite{KorMar}. Furthermore, the compression-equivocation rate region depends on the joint distribution $p_{ABE}$ only via its marginals $p_{AB}$ and $p_{AE}$. Hence, physical degradation and stochastic degradation are equivalent in this scenario.

\begin{cor}\label{c:lessnoisy}
For uncoded side information at Bob, if Bob has \textit{less noisy} side information than Eve, then an $(R_A,\Delta)$ pair is achievable if and only if
\begin{eqnarray}
  R_A &\geq& H(A|B), {~\rm and} \\
  \Delta & \leq & I(A;B)-I(A;E).
\end{eqnarray}
\end{cor}

\begin{proof}
Achievability follows simply by letting $U$ be constant in Corollary \ref{c:uncodedSI}. For the converse, consider any $U$ with the joint distribution $p(u,a,b,e) = p(a,b,e)p(u|a)$. We have
\begin{align}
&[I(A;B) - I(A;E)] - [I(A;B|U) - I(A;E|U)] \nonumber \\
& = [I(A;B) - I(A;E)]  \nonumber \\
& -  [I(A,U;B) -  I(B;U) - I(A,U;E) + I(E;U)] \\
&= I(B;U|E) - I(E;U|B)\\
&= I(B;U) - I(E;U) \geq 0,
\end{align}
where the last inequality is due to the less noisy assumption. 
\end{proof}

Corollary \ref{c:lessnoisy} for the special case of physically degraded side information at Eve is given in \cite{Merhav} as well. The following corollary, which we state without proof, gives a condition under which no positive equivocation can be achieved.

\begin{cor}\label{c:phydeg}
If Bob's side information is a stochastically degraded version of Eve's side information, then no positive equivocation rate is achievable, and $\Delta=0$.
\end{cor}



We use the following simple example (suggested in \cite{Ramc}) to illustrate some of our results. Let the original source sequence $A^N=(A_1, \ldots, A_N)$ available to Alice
be an i.i.d. binary sequence of $A_i \sim Bernoulli(1/2)$ random variables. The observation of Bob $B^N=(B_1, \ldots, B_N)$ is generated by independently erasing each element of the $A^N$ sequence with probability $p_B$, that is, $B_i =A_i$ with probability $1-p_B$, and $B_i=e$ with probability $p_B$. Similarly, the observation $E^N=(E_1, \ldots, E_N)$ of the eavesdropper Eve is an independent erased version of $A^N$. We have $E_i =A_i$ with probability $1-p_E$, and $E_i=e$ with probability $p_E$.

For $p_E > p_B$, the side information of Eve is a stochastically degraded version of the side information of Bob. Using Corollary \ref{c:lessnoisy}, we know that a constant $U$ is optimal. Then, the optimal equivocation is $\Delta = I(A;B) - I(A;E) = (1 - p_B) - (1-p_E) = p_E - p_B$.

When $p_B \geq p_E$, then $B^N$ is a stochastically degraded version of $E^N$. From Corollary \ref{c:phydeg}, we get $\Delta=0$.

\section{Side information available to Alice } \label{s:side_info}

In this section, we consider various cases in which Alice also has access to the side information available to Bob and/or Eve. We know from the Slepian-Wolf source coding that, the availability of Bob's side information at Alice does not help in terms of compression rates. However, as shown in \cite{Ramc} via a simple example, in the secure compression setup, the availability of $B^N$ at Alice potentially enables higher equivocation rates at the eavesdropper. In the following theorem, we characterize the compression-equivocation rate regions for various side information scenarios at Alice.

\begin{thm}\label{t:SI}
Consider secure source compression for uncoded side information at Bob as illustrated in Fig. \ref{f:uncodedSI}. An $(R_A,\Delta)$ pair is achievable if and only if
\begin{eqnarray}
  R_A &\geq& H(A|B),  {~\rm and} \\
  \Delta & \leq & \max \{ I(A;B|U) - I(A;E|U)\},
\end{eqnarray}
where we maximize over auxiliary random variables $U$ such that the joint distribution $p(u,a,b,e)$ is given in the following table depending on which switches are closed:

\vspace{.2in}
\centering
\begin{tabular}{|c|c|}
  \hline
  Closed Switches & $p(u,a,b,e)$ \\
  \hline
  $S_B$ & $p(a,b,e)p(u|a,b)$ \\
  $S_E$ & $p(a,b,e)p(u|a,e)$ \\
  $S_B$ and $S_E$ & $p(a,b,e)p(u|a,b,e)$ \\
  \hline
\end{tabular}
\end{thm}
\vspace{.2in}

In the case when only the switch $S_E$ is closed, the rate region can be explicitly given as follows.
\begin{eqnarray}
  R_A \geq H(A|B) \mbox{  and  } \Delta \leq I(A;B|E). \label{e:SE_closed1}
\end{eqnarray}
\begin{proof}
The proof resembles Theorem \ref{t:codedSI}, and will not be included due to space limitations.
\end{proof}

Note that the availability of either or both of the side information sequences at the transmitter enlarges the space of the auxiliary random variables $U$ and potentially results in a higher equivocation rate at the eavesdropper. To illustrate this, consider the random erasure side information example in Section \ref{s:no_enc_SI}. Suppose that the observation of Bob $B^N$ is available to Alice as well. Alice can transmit only the erased bits of Bob, hence leaking the least amount of information to Eve. As stated in \cite{Ramc}, it is possible to show that the optimal auxiliary random variable $U$ satisfies $U=A$ when there is an erasure at Bob, and $U$ is constant otherwise. The optimal equivocation rate in this case\footnote{There is a typo in the leakage rate of $1-p_Yp_Z$ reported in \cite{Ramc}. It should have been $1-p_Z-p_Yp_Z$. } is $\Delta = p_E(1-p_B)$. Note that this equivocation is strictly larger than the one without side information. Furthermore, even if Bob's side information is a stochastically degraded version of Eve's, i.e., $p_B > p_E$, we are still able to achieve a non-zero equivocation rate if this side information can be provided to Alice as well.


When only the observation of Eve, $E^N$ is available to Alice, from (\ref{e:SE_closed1}) the optimal equivocation rate is given by $I(A;B|E)$. In the erasure example, the optimal equivocation rate is found to be $\Delta = p_E(1-p_B)$, which is the same as in the case when only switch $S_B$ is closed. We observe that, for this specific example of erased observations at Bob and Eve, the benefit of having either Bob's or Eve's side information to Alice is the same. For this example, it is also possible to show that, even when both observation sequences are available to Alice, the optimal equivocation rate is still $\Delta = p_E(1-p_B)$.

While there is no difference between physically or stochastically degraded observations when both switches are open, this is no longer true when we consider side information at Alice. In the following corollary, we show that for a physically degraded observation at Eve, the availability of $E^N$ to Alice does not help. This is in contrast to stochastically degraded side information $E^N$ whose availability at Alice would potentially increase the equivocation rate as seen in the example above.

\begin{cor}
If the observation of Eve is a physically degraded version of Bob's side information, i.e., $A-B-E$ form a Markov chain, then providing this observation to Alice would not improve the equivocation rate.
\end{cor}

%

%

\section{Conclusion}
We have considered secure lossless compression in the presence of an eavesdropper with correlated side information. We have shown that secure communication can be enabled by another agent who has its own correlated side information and a secure link to the legitimate receiver. We have studied scenarios under which secure compression codebooks are identical to Slepian-Wolf codebooks. We have also characterized the compression-equivocation rate regions considering availability of side information at the encoder. We have shown that, while it is useless in the pure lossless compression setup, side information at the encoder may help to increase the equivocation rate in secure compression model.

\appendices
\section{Proof of Theorem \ref{t:codedSI}}\label{App1}

\emph{Inner bound:} We fix $p(u|a)$ and $p(v|c)$ satisfying the conditions in the theorem. Then we generate $2^{N(I(A;U)+\epsilon_1)}$ independent codewords of length $N$, $U^N(w_1)$, $w_1 \in \{1,\ldots,2^{N(I(A;U)+\epsilon_1)} \}$, with distribution $\prod_{i=1}^N p(u_i)$.  We randomly bin all $U^N(w_1)$ sequences into $2^{N(I(A;U|V)+\epsilon_2)}$ bins, calling them the auxiliary bins. For each codeword $U^N(w_1)$, we denote the corresponding auxiliary bin index as $a(w_1)$. On the other hand, we randomly bin all $A^N$ sequences into $2^{N(H(A|V,U)+\epsilon_3)}$ bins, calling them the source bins, and denote the corresponding bin index as $s(A^N)$. We also generate $2^{N(I(C;V)+\epsilon_4)}$  independent codewords $V^N(w_2)$ of length $N$, $w_2 \in \{1,\ldots,2^{N(I(C;V)+\epsilon_4)} \}$, with distribution $\prod_{i=1}^N p(v_i)$.

For each typical outcome of $A^N$, Alice finds a jointly typical $U^N(w_1)$. Then she reveals $a(w_1)$, the auxiliary bin index of $U^N(w_1)$, and $s(A^N)$, the source bin index of $A^N$, to both Bob and Eve, that is, the encoding function $f_A$ of Alice is composed of the pair $(a(w_1), s(A^N))$. Using standard techniques, it is possible to show that we have such a unique index pair with high probability.

The helper, Charlie, observes the outcome of its source $C^N$, finds a jointly typical $V^N$ with $C^N$, and sends the index $w_2$ of $V^N$ over the private channel to Bob. With high probability $C^N$ will be a typical outcome, and there will be a unique $V^N(w_2)$ that is jointly typical with $C^N$. Bob, having access to $V^N(w_2)$ and the auxiliary bin index $a(w_1)$, can find the jointly typical $U^N(w_1)$ correctly with high probability. Then using $V^N(w_2), U^N(w_1)$ and the source bin index $s(A^N)$, Bob can reliably decode the source sequence $A^N$. Letting $\epsilon_i \rightarrow 0$ for $i=1,2,3$ and
$4$, we can make the total communication rate of Alice arbitrarily close to $I(A;U|V) + H(A|U,V) = H(A|V)$, while having an error probability less than $\epsilon$ for sufficiently large $N$.

The equivocation rate for this scheme can be found as
\begin{align}
& \frac{1}{N}H(A^N|a(w_1), s(A^N), E^N) \nonumber \\
 &= \frac{1}{N} \left[ H(A^N) - I(A^N; a(w_1), s(A^N), E^N) \right]  \nonumber \\
&= \frac{1}{N} \left[ H(A^N) - I(A^N; a(w_1), E^N) \right. \nonumber \\
 &~~~~~~ - \left. I(A^N; s(A^N) | E^N, a(w_1)) \right]  \nonumber \\
 &\geq \frac{1}{N} \left[H(A^N) -  I(A^N; U^N, E^N) - H(s(A^N)) \right]  \label{ach1} \\
 &= H(A| U, E) - H(A| V, U) -\epsilon_3  \label{ach2} \\
 &= I(A; V |U) - I(A; E |U) -\epsilon_3,  \nonumber
\end{align}
where (\ref{ach1}) follows form the data processing inequality; and (\ref{ach2}) follows form the fact that $s(A^N)$ is a random variable over a set of size $ 2^{N (H(A| V, U) + \epsilon_3)}$. 

Finally, we also have
\begin{align}
& \frac{1}{N}H(A^N|a(w_1), s(A^N), E^N) \nonumber \\
 &= \frac{1}{N} \left[ H(A^N|E^N) - I(A^N; a(w_1), s(A^N)| E^N) \right]  \nonumber \\
&\geq H(A|E) - \frac{1}{N} H(a(w_1), s(A^N)) \\
 &\geq H(A|E) - R_A.  \label{n_ach1}
\end{align}

\emph{Outer bound:} Let $J \triangleq f_A(A^N)$ and $K \triangleq f_C(C^N)$. From Fano's inequality, we have $H(A^N | J, K)\leq N \delta (P_e^N),$ where $\delta (x)$ is a non-negative function with $\lim_{x \rightarrow 0} \delta(x) =0$.

Define $U_i\triangleq (J,A^{i-1},E^{i-1})$ and $V_i \triangleq (K,C^{i-1})$.  Note that both $U_i-A_i-(B_i,E_i)$ and $V_i-C_i-(A_i,E_i)$ form Markov chains. Then, we have the following chain of inequalities:
\begin{align}
N R_C \geq & H(K) \geq I(C^N; K) = \sum_{i=1}^N I(C_i;K,C^{i-1}) \label{ee:c11} \\
 = & \sum_{i=1}^N I(C_i;V_i), \nonumber
\end{align}
where (\ref{ee:c11}) follows from the chain rule of mutual information and the memoryless assumption on $C_i$. We also have
\begin{align}
N R_A \geq& H(J) \geq H(J|K) \nonumber \\
 = & H(A^N, J| K) - H(A^N|J,K) \nonumber \\
 \geq & H(A^N|K) - N\epsilon \label{ee:c21} \\
 = & \sum_{i=1}^N H(A_i|K,A^{i-1}) - N\epsilon \nonumber \\
 \geq & \sum_{i=1}^N H(A_i|K,A^{i-1}, C^{i-1}) - N\epsilon  \label{ee:c22} \\
 = & \sum_{i=1}^N H(A_i|K,C^{i-1}) - N\epsilon  \label{ee:c23} \\
 = & \sum_{i=1}^N H(A_i|V_i) - N\epsilon,  \nonumber
\end{align}
where (\ref{ee:c21}) follows from Fano's inequality and nonnegativity of entropy; (\ref{ee:c22}) follows as $A_i-(K,A^{i-1})-C^{i-1}$ form a Markov chain; and (\ref{ee:c23}) follows as $A_i-(K,C^{i-1})-A^{i-1}$ form a Markov chain. 

Finally, we can also obtain
\begin{align}
& H(A^N| J, E^N) = H(A^N|J) - I(A^N;E^N|J) \nonumber \\
& = H(A^N| J,K) + I(A^N; K|J) - I(A^N;E^N|J) \nonumber \\
& = \sum_{i=1}^N I(A_i;K|J,A^{i-1}) - H(E_i|J,E^{i-1}) \nonumber \\
& ~~~~~~~~~~~~~~~~  + H(E^N | A^N,J) + N\epsilon \label{eq_1_3} \\
&\leq \sum_{i=1}^N I(A_i;K|J,A^{i-1}, E^{i-1}) - H(E_i|J,E^{i-1}, A^{i-1}) \nonumber \\
&~~~~~~~~~~~~~~~~~  + H(E^N | A^N) + N\epsilon \label{eq_1_4} 
\end{align}
\begin{align}
&\leq \sum_{i=1}^N  \left[ I(A_i;K,C^{i-1}|J,A^{i-1}, E^{i-1})  \right. \nonumber \\
&~~~~~ - \left. H(E_i|J,E^{i-1}, A^{i-1}) + H(E_i | A_i) \right] + N\epsilon \label{eq_1_4b} \\
&= \sum_{i=1}^N \left[ I(A_i;V_i|U_i) - H(E_i|U_i) + H(E_i|A_i)  \right]+ N\epsilon \label{eq_1_5} \\
&= \sum_{i=1}^N \left[ I(A_i;V_i|U_i) - I(A_i; E_i|U_i) \right]+ N\epsilon \label{eq_1_5b}
\end{align}
where (\ref{eq_1_3}) follows from the Fano's inequality and the chain rule of mutual information; (\ref{eq_1_4}) follows from the memoryless property of the source and the side information sequences, and the fact that conditioning reduces entropy; (\ref{eq_1_4b}) follows from the chain rule and non-negativity of mutual information; (\ref{eq_1_5}) follows from the definitions of $V_i$ and $U_i$ given above and the fact that conditioning reduces entropy; (\ref{eq_1_5b}) follows since $U_i-A_i-E_i$. 

We define an independent random variable $Q$ uniformly distributed over the set $\{1,2,\ldots,N\}$, and  $A=A_Q$, $E=E_Q$, $V=(V_Q, Q)$, and $U=(U_Q,Q)$. Then from the usual techniques, (\ref{rateA})-(\ref{rateD}) follow while $V-C-(A,E)$ and $U-A-(C,E)$ are Markov chains. Finally, we also have
\begin{align}
\frac{1}{N} H(A^N|E^N) &\leq \frac{1}{N} H(A^N, J|E^N)  \nonumber\\
& = \frac{1}{N} \big[ H(J|E^N) + H(A^N|E^N, J)  \big] \nonumber\\
& \leq \frac{H(J)}{N} + \Delta \leq R_A + \Delta. \nonumber
\end{align}

\end{document}